**Response to the comment of K.W. Edmonds *et al.* on the article "Controlling the Curie temperature in (Ga,Mn)As through location of the Fermi level within the impurity band"**

M. Dobrowolska[1]*, X. Liu[1], J. K. Furdyna[1], M. Berciu[2], K. M. Yu[3] and W. Walukiewicz[3]

Although we seriously disagree with many of the points raised in the comment by Edmonds *et al.*[1], we feel that it is valuable and timely, since comparison of this comment and our paper[2] serves to underscore an important property of the ferromagnetic semiconductor (Ga,Mn)As in thin film form. In an earlier publication Yu *et al.*[3] have shown that when the thickness $d$ of a (Ga,Mn)As film is ultra-thin (typically for $d < 50$ nm), the film will manifest very different stoichiometric and magnetic behavior from bulk (Ga,Mn)As (as represented by specimens with $d > 65$ nm). As will be seen below, our disagreement with the Comment of Edmonds *et al.*[1] can largely (although not entirely) be traced to the differences between ultra-thin (Ga,Mn)As films and thicker bulk-like material.

The Comment raises several points, which we will address in the order in which they were presented.

Item (i) of the Comment:

The authors raise the point that the calculations in Jungwirth *et al.*[4] are not based on a "valence band model," as was stated in our paper. We feel that this distinction is somewhat semantic. It is a standard practice to broadly divide theories in this field into *impurity band (IB)* or *valence band (VB)* models, depending on whether they predict the Fermi energy $E_F$ to be above or below the top of the valence band, i.e. whether $E_F$ resides among impurity-like states or valence-band-like states. The overall result of the calculations of Ref. 4 is that the impurity band is completely merged with the valence band; and that the ferromagnetic coupling in (Ga,Mn)As is mediated by itinerant holes in the valence band that are, to a good extent, similar to free valence band holes. This was our reason for including it in the VB-model category. Of course there are significant differences in the level of detail between various VB models. One common feature, however, is that all VB models predict a *monotonic increase* of $T_C$ with both Mn and hole concentrations. We chose to compare our results with the prediction of Ref. 4 because the model used in that reference is more detailed and quantitative than others in the same category. The main conclusion of our paper is that our samples do not show the monotonic increase of $T_C$ with $p$ and $x_{eff}$, and are thus inconsistent with the generic predictions of any VB model.

In item (i) of the Comment its authors also state that our paper does not mention the experimental results presented in Ref. 4, which (as shown in Fig 1 of the Comment) disagree with our non-monotonic behavior of $T_C$ as a function of compensation level. However, as argued below in discussing items (iii) and (iv) of the Comment, the experimental results in Ref. 4 are not really relevant to our experiments (nor in fact to the theory presented in Ref. 4).

Item (ii):

In this item the authors state that "the claimed agreement between [our data in Ref. 2] and the impurity band model is solely based on a cartoon," and they raise the question why we did not present, or refer to, values of $T_C$ calculated from a quantitative impurity band theory. We disagree with the statement that the agreement of our results with the IB model is based "solely on a cartoon". In addition to the non-monotonic behavior of $T_C$ vs $p$ and $x_{eff}$, which is inconsistent with all VB models, we also show, for example, the results of magnetic circular dichroism (MCD) measurements *on the same set of samples*. These confirm that $E_F$ lies above the VB states, consistent with IB models as defined above. Many additional arguments are also listed in our paper and in the Supplementary Material to support this picture. Furthermore, there exists a large body of other experimental work which disagree with "VB models," one of the most recent being that of Ohya *et al.*[5], where the authors find little or no coupling between VB states and the Mn spins, as well as other behaviors which are also inconsistent with any generic VB model, but qualitatively consistent with IB models.

The absence of detailed microscopic theories to explain how the IB could "still avoid mixing and overlapping with the VB band at $10^{20}$ - $10^{21}$ cm$^{-3}$ doping levels" is simply due to the difficulty of accurately dealing with disorder and electron-electron interactions in such systems. However, based on a very simple one-band model, a recent theoretical study[6] shows that, in the absence of electron-electron interactions, positional disorder of non-magnetic dopants which occupy $x$ percent of sites results in an IB-like feature in the density of states (DOS) up to $x =$ 10% and above. Depending on various parameters, the IB may be separated, or may be in contact with its parent band; in all cases, however, the eigenstates in the IB have IB-like character, i.e. they primarily reside at the impurity sites. Simple methods of dealing with disorder, like the coherent potential approximation, fail to capture the IB nature of these wavefunctions. Furthermore, the addition of electron-electron interactions is likely to push the IB even further from its parent band. After all, correlations can open gaps even in clean systems. A long time ago Mott stated, in discussing disordered semiconductors, that "At the metal-insulator transition the impurity band is thought to be separate from the conduction band, and to merge with it for concentrations about ten times higher"[7]. The lack of approximation-free numerical simulations of this complicated problem is due to the difficulty of the task; but it certainly does not mean that Mott's ideas about the IB picture are wrong.

Item (iii):

The authors of the comment state that our maximum value of $T_C$ is only 90 K, and therefore our results have not demonstrably led to high values of $T_C$. While we do say that having a correct model for ferromagnetism of (Ga,Mn)As would be beneficial in formulating strategies for optimizing this material, including its $T_C$, the main purpose of our paper was to understand the properties of intrinsic (Ga,Mn)As generally, i.e., in a wide range of compositions, compensations, and transport properties, from insulating to metallic, rather than just metallic samples with highest values of $T_C$. Here we should emphasize that our samples are 100 nm thick, and thus are representative of bulk (Ga,Mn)As. As was shown by Ku *et al.*[8], the sample thickness limits the maximum obtainable Curie temperature. More specifically, samples with thicknesses $d \leq 50$nm show much higher values of $T_C$ than samples with d $\geq 65$ nm despite the fact that they are grown under identical conditions. The high values of $T_C \sim 188$K cited in the Comment were only reached on very thin films ($d = 25$ nm[9] and $d = 23$ nm[10]). According to Ref. 3 such samples are greatly affected by surface and/or interfacial effects, and thus are not representative of bulk (Ga,Mn)As.

<u>Item (iv):</u>

Referring to our use of ion channeling methods for determining hole and Mn concentrations, in item (iv) the authors of the Comment state that in their opinion such methods do not provide a direct measurement of either of these parameters, and that they also neglect other important contributions, such as compensating defects (e.g., As antisites) and sample inhomogeneities. This issue can be addressed by re-plotting the graph of $T_C/x_{eff}$ vs. $p/x_{eff}$ shown in Fig. 1 of our paper using values of $p$ established by an independent method, other than channeling. We have therefore plotted $T_C/x_{eff}$ vs. $p/x_{eff}$ for a series of (Ga,Mn)As samples using values of $p$ obtained by electrochemical capacitance voltage profiling (ECV) measurements, as shown in Fig. 1 of this response. The data in the new plot are obtained as follows. The black dots are obtained by using the ECV values of $p$ (as well as the corresponding values of $T_C$ and $x_{eff}$) taken from the paper of Wojtowicz et al.[11] The blue squares are data points taken from the article of Cho et al.[12] And the points shown in red correspond to samples studied in our original paper that were still available, in which the value of $p$ has now been re-measured using the ECV technique. The new plot, obtained by using only ECV values for $p$, remains highly non-monotonic, confirming the conclusions of our original article in Nature Materials.

In item (iv) the authors also state that, based on Fig. 12 from Ref. 4 and Fig.1 of the Comment, their experimental results disagree with ours in that they do not show the collapse of $T_C$ at low compensations. In our opinion the experimental results of Ref. 4, while very interesting, are of little relevance to the discussion of intrinsic properties of (Ga,Mn)As, since they were obtained on ultra-thin films (25 and 50 nm). Additionally, the methods used to determine the key parameters of these specimens are questionable, for the following reasons. First, the total Mn content, $x_{tot}$, was determined by SIMS calibration measurements taken on films of 1 micrometer thickness. Although the control samples and the ultrathin films used in Ref. 4 were grown under the same MBE conditions, the incorporation and the site distribution of Mn atoms depends critically on sample thickness[3], so the extrapolation from 1000 to 25 nm is highly questionable. Second, the hole concentration $p$ in Ref. 4 is determined by *ad hoc* generated arguments and fitting to an untested formula for the magnetic-field-dependent anomalous Hall effect. Third, this hole concentration is also used together with the total Mn concentration established by SIMS to obtain the value of $x_{eff}$. Apart from the error inherent in $p$ and in $x_{tot}$, as discussed above, in calculating $x_{eff}$ the authors of Ref. 4 ignore the concentration of random Mn precipitates which, as shown by Yu et al.,[3] grows rapidly with decreasing thickness, and is entirely different in bulk and in ultrathin (Ga,Mn)As. Finally, in calculating $x_{eff}$, Ref. 4 ignores all sources of compensation other than $Mn_I$. As a result, the values of $x_{eff}$ so obtained carry a very large systematic error. This may be the reason that points in Fig 12 of Ref. 4 and in Figs. 1 and 2 of the Comment show values of $p/x_{eff}$ larger than unity, something that is completely unphysical.

Thus the principal difference between our results and that of Ref. 4 (and also Ref. 13 to which the comment refers) is the thickness of the samples used. While our samples are 100 nm thick and can be treated as bulk[3], the samples in both references are considerably thinner (25 and 50 nm in Ref. 13), and thus should not be compared with theories developed for intrinsic (Ga,Mn)As, for the following reasons. The authors claim that their ultrathin annealed films are perfect GaAs crystals doped with only substitutional Mn, while in reality the small thickness of the samples invalidates some of the arguments that lie at the basis of the analysis of the experimental results used in Ref. 4. Apart from the issues related to random Mn precipitates raised in the preceding paragraph, the films have an interface with low-temperature-grown GaAs on one side and the free surface on the other side. Both the interface and the free surface have Fermi energy pinned by the native defects in the band gap, and both can serve as practically infinite sources of defects. In addition, the electric fields in the interface/surface depletion regions will further enhance diffusion of charged native defects. Thus defects such as native interstitials and/or vacancies can easily diffuse into such thin layers, significantly changing their properties, especially in samples 25 nm thick that were annealed for as long 150 hrs at 190 C.

We emphasize that we do not question the quality of the materials used in Ref. 4. Samples such as those indeed lead to highest values of $T_C$ seen to date, and are thus very important; but the data obtained on such samples must be analyzed in terms of the balance between the surface and the bulk, rather than treated as pure bulk. On the other hand, even though our thicker films have lower values of $T_C$, they provide a closer approximation to the uniform bulk material, and are thus more appropriate for a study of the origin of ferromagnetism intrinsic to (Ga,Mn)As.

[1]Department of Physics, University of Notre Dame, Notre Dame, Indiana 46556, USA;
[2]Department of Physics and Astronomy, University of British Columbia, Vancouver, British Columbia, Canada;
[3]Materials Science Division, Lawrence Berkeley National Laboratory, Berkeley, California 94720, USA.
*e-mail: mdobrowo@nd.edu.

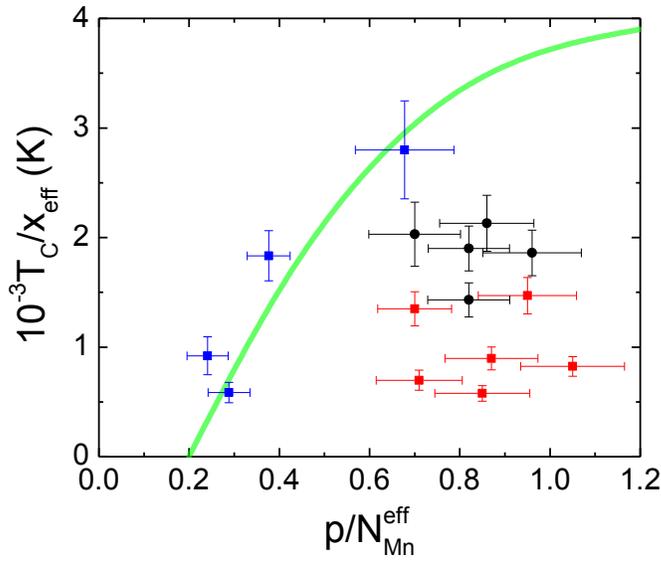

Figure 1. Comparison between experimental data and theoretical calculations based on the valence band model of Jungwirth *et al.*, Ref. 4. The calculations and the data shown in the figure are plotted as $T_C/x_{eff}$ vs $p/N_{Mn}^{eff}$ (where $p/N_{Mn}^{eff}$ is the ratio of the hole concentration to the concentration of effective Mn moments $N_{Mn}^{eff}$, with $N_{Mn}^{eff} = 4x_{eff}/a^3$). The experimental data shown in the figure are taken from Ref. 11 (black dots), Ref. 12 (blue squares), and from our original paper[2], where the values of $p$ have now been re-measured using the ECV technique (red squares). The error bars in the figure account for 5% accuracy in determining $x_{eff}$ and 2% accuracy in determining the hole concentration from ECV experiments.